\documentclass{sig-alternate}

\usepackage[utf8]{inputenc}
\usepackage{graphicx} 
\usepackage{graphics} 
\usepackage{wrapfig}
\usepackage{color}
\usepackage[table]{xcolor}
\usepackage{hyperref}
\newcommand{\aspas}[1]{{``#1''}}
\usepackage{array}
\usepackage{enumerate}
\usepackage{setspace}
\usepackage{soul}
\usepackage{changepage}

\newcolumntype{P}[1]{>{\centering\arraybackslash}p{#1}}

\newcommand{\frase}[2]{%
\noindent
\begin{center}
\begin{adjustwidth}{0.8cm}{0.8cm}
\aspas{\em #1}\\[0.1cm]
\hspace*{\fill}{({\em#2})}
\vspace{5pt}
\end{adjustwidth}
\end{center}
}

\begin{document}
	
	\title{Does Technical Debt Lead to the Rejection of \\Pull Requests?}
	
	%
	%
	%
	%
	
	\numberofauthors{3} 
	%
	\author{
		%
		%
		\alignauthor
		Marcelino Campos Oliveira Silva\\
		\affaddr{{Chemtech - A Siemens Business}}\\
		\affaddr{Belo Horizonte, Brazil}\\
		\email{\hspace*{-10pt}marcelino.silva@siemens.com}
		\alignauthor
		Marco Tulio Valente\\
		\affaddr{Departamento de Ci\^{e}ncia da Computa\c{c}\~{a}o}\\
		\affaddr{Universidade Federal de Minas Gerais}\\
		\affaddr{Belo Horizonte, Brazil}\\
		\email{mtov@dcc.ufmg.br}
		\alignauthor Ricardo Terra\\
		\affaddr{Departamento de Ci\^{e}ncia da Computa\c{c}\~{a}o}\\
		\affaddr{Universidade Federal de Lavras}\\
		\affaddr{Lavras, Brazil}\\
		\email{terra@dcc.ufla.br}
	}
	
	\maketitle

\begin{abstract}
Technical Debt is a term used to classify non-optimal solutions during software development. 
These solutions cause several maintenance problems and hence they should be avoided or at least documented.
Although there are a considered number of studies that focus on the identification of Technical Debt, we focus on the identification of Technical Debt in pull requests. 
Specifically, we conduct an investigation to reveal the different types of Technical Debt that can lead to the rejection of pull requests. 
From the analysis of 1,722 pull requests, we classify Technical Debt in seven categories namely design, documentation, test, build, project convention, performance, or security debt. 
Our results indicate that the most common category of Technical Debt is design with 39.34\%, followed by test with 23.70\% and project convention with 15.64\%. 
We also note that the
type of Technical Debt influences on the size of push request discussions, e.g., security and project convention debts instigate more discussion than the other types.
\end{abstract}


\section{Introduction}
Software developers usually face challenges to build complex systems in a short deadline. More relevant, in order to accomplish such deadlines, they may implement their tasks using sub-optimal technical approaches, which increases the costs of the maintenance in the future. When this situation ascends, the project acquires Technical Debt~\cite{4}. 

In this paper, we focus on Technical Debt through code submission via pull requests. Our goal, thereupon, is to investigate  what are the most common Technical Debts that cause rejection of pull requests.

Such investigation provides open source developers with identification of the most frequent types of Technical Debts,
which leads
at least to the following contributions:
(i)~the identification of a high occurrence of one type of problem can alert both reviewers and developers;
(ii)~developers can produce a checklist of the most common Technical Debts before submitting pull requests;
and 
(iii)~reviewers can better evaluate pull requests when they are 
aware of some types of problems, e.g., a reviewer may approve a code that successfully solves a particular bug, but in return, brings problems in the management of memory, which will only be realized in the future. 

This paper is organized as follows. Section~\ref{sec:background} provides background information about
pull-based development
and technical debt. Section~\ref{sec:studydesign} describes the methodology of our study, stating
four research questions and presenting our dataset.
Section~\ref{sec:results} answers and discusses each research question.
Finally,
Section~\ref{sec:threats} states threats to validity
and Section~\ref{sec:conclusion} concludes.

\begin{figure*}[!t]
\vspace{10pt}
\begin{center}
\includegraphics[width=0.8\textwidth]{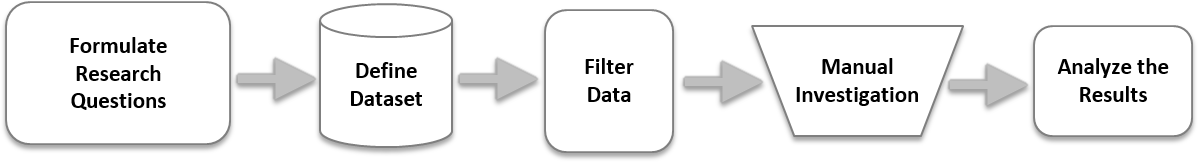}
\caption{Study Design Overview}
\label{fig:design}
\end{center}
\end{figure*}

\section{Background}
\label{sec:background}

\subsection{Distributed Software Development}
\label{sec:dsd}
Distributed software development allows developers to work on a software project even being geographically dispersed. There are a bunch of integration techniques that help to produce software in a distribute way, but a particular one has become very popular nowadays: pull-based development~\cite{8}.

The pull-based development process has been widely used and is quickly becoming the default one. 
Large source code hosting sites (e.g., GitHub, BitBucket, and Gitorious) support pull-based development, totalizing more than a million projects, which indicates its current relevance. 
In pull-based development, a developer pushes changes to a repository and describes the underlying changes, e.g., new features, bug fixes, improvements, etc.
Soon after, developers submit a pull request, interested parties evaluate it in order
to accept or reject its integration to the project.

In GitHub\footnote{\url{https://developer.github.com/guides/working-with-comments/}}, for instance, contributors and reviewers can interact through three kinds of comment views: (i)~{comments on the pull request itself}, (ii)~{comments on a specific line within the pull request}, and (iii)~{comments on a specific commit within the pull request}. 
The acceptance or rejection of a pull request can trigger many interactions between the contributors and reviewers. 
This process is not trivial since it involves both technical and social communication aspects~\cite{15}.

Pull request evaluation varies from a project to another. Gousios et al.~\cite{9} investigated quantitatively what factors are behind the acceptance in pull-based development. They report that all projects have a similar process, but the dominating factors (stickiness of project area and social distance, respectively) are massively different. 
On one hand, this difference suggests that there is no common process for evaluating contributions in pull-based development. On the other hand, there are cases in which the owner relies on bots to inspect and reject pull requests that do not comply with contributor license agreement.

Bacchelli and Bird~\cite{2} investigate a specific technique used in pull-based development named \aspas{modern} code review. 
Although it is very similar to the traditional code review,
\aspas{modern} code review determines {\em a priori} the process for accepting a contribution
and requires sign-off by a specific number of collaborators.
They argue that, although the main purpose of \aspas{modern} code review is to find defects, in practice 
the benefits of knowledge transfer and team awareness counterpoise these purposes.

\subsection{Technical Debt}
\label{sec:technical_debt}

In our study, we rely on the terminology coined by Fowler~\cite{7} which defines that
Technical Debt occurs when developers adopt an un-optimal or sub-optimal technical decision. Therefore, solutions that do not solve the whole problem are not considered in our study. 
Alves et al.~\cite{1}, for example, consider bugs and unimplemented requirements as types of Technical Debt. In contrast, we do not consider them as Technical Debt since they do not solve the original problem they have been proposed to. 
We also rely on the definition of Curtis et al.~\cite{5} that state that Technical Debt should be distinguished from defects or failures. 
Failures during testing or operation can be symptoms of Technical Debt, but most structural issues related to Technical Debt may not lead to test or operational failures. 
Instead, they may cause a less efficient use, less scalable, more difficult to maintain, or more vulnerable system. 
In essence, Technical Debt emerges from poor design quality and affects business as well as IT costs and business risks. 

Similarly to our study, there are studies that focus on the identification of Technical Debt. For example, Santos et al.~\cite{16}, Letouzey et al.~\cite{11} and Marinescu~\cite{13} use metrics based on cohesion, coupling, code duplication, lack of comments, coding rules violation, potential bugs, and the absence of unit tests to reveal the existence of Technical~Debt.

There also are other studies that identify Technical Debt in a restricted way. 
Zazworka et al.~\cite{17} focus on a particular kind of design debt, namely God Classes.
Fontana et al.~\cite{6} investigate design Technical Debts that appear in the form of bad smells. 
Potdar and Shihab~\cite{14}, and Maldonado and Shihab~\cite{12} investigate the detection of Technical Debt by manually inspecting code comments. 

Zazworka et al.~\cite{4} conduct an experiment to compare the efficiency of automated tools in comparison with human elicitation, regarding the detection of Technical Debt.
They report that there is a small overlap between the two approaches and hence the best solution is to combine them. 
They conclude that automated tools are more efficient in finding defect Technical Debt, whereas developers can realize more abstract categories of Technical Debt.

\section{Study Design}
\label{sec:studydesign}

As illustrated in Figure~\ref{fig:design}, we define the following protocol to conduct our study:
\begin{enumerate}[~~~(a)]
\item Define research goals and questions, 
\item Define dataset and resources, 
\item Define study data filtering procedure and exclusion criteria,
\item Conduct manual inspection, and 
\item Analyze the results.
\end{enumerate}

The next subsections describe in details each aforementioned step.

\begin{table*}[ht]
\centering
\caption{Target Projects}
  \begin{tabular}{lp{7cm}p{4.5cm}cc}
\hline
{\bf System}	& {\bf URL} 	& {\bf Domain}	& {\bf LOC}	& {\bf PR's}\\
\hline
Elastic Search &	https://github.com/elastic/elasticsearch	& Full-text search engine &	428,550 & 	6,115\\
IntelliJ &	https://github.com/JetBrains/intellij-community	 & Programming IDE & 	504,505	 & 323\\
Iosched	 & https://github.com/google/iosched & 	Conference Application	 & 41,666	 & 72\\
Picasso & 	https://github.com/square/picasso & 	Image downloading library	 & 9,238	 & 344\\
Retrofit & 	https://github.com/square/retrofit	 & HTTP client for Android  & 	6,013	 & 523\\
Storm	 & https://github.com/apache/storm/ & 	Distributed real-time system & 	46,306	 & 262\\
\hline 
{\bf Total}		&&&&		{\bf 7,639}\\
 \hline 
  \end{tabular}

  \label{tb:projects}  
\end{table*}

\subsection{Research Questions}

This study aims to understand how the rejection of pull requests relates to Technical Debt. 
Using the Goal-Question-Metric format proposed by Basili et al.~\cite{3}, 
we defined the following research questions:\\

\newcommand{\rqi}{{\noindent\centering\bf\em RQ~\#1: How often pull requests are rejected?}}
\rqi\\[-0.2cm]

This question investigates the percentage of rejection in pull requests.\\

\newcommand{\rqii}{{\noindent\centering\bf\em RQ~\#2: Does Technical Debt lead to the rejection of pull requests?}} 
\rqii\\[-0.2cm]

This question investigates whether reviewers and contributors list Technical Debt as the main reason to reject pull requests.\\

\newcommand{\rqiii}{{\noindent\centering\bf\em RQ~\#3: What are the most common types of Technical Debt that lead to the rejection of pull requests?}} 
\rqiii\\[-0.2cm]

Assuming a positive answer to RQ~\#2, this research question goes further. Although Technical Debt is a general term, some studies reported different types of Technical Debt~\cite{1}. 
Nevertheless, the types of Technical Debts that occur in pull requests remain unknown.

\newcommand{\rqiv}{{\noindent\centering\bf\em RQ~\#4: 
Which types of Technical Debt spur more discussion? 
}}
\rqiv\\[-0.2cm]

This research question, which also assumes a positive answer to RQ~\#2,
verifies whether a particular type of Technical Debt instigates more discussion
than the others.
Therefore, we investigate whether contributors discuss more to reject and close a pull request
incurring in a specific type of debt, rather than others.

\subsection{Dataset}
Our study relies on six open source projects, namely IntelliJ, Elastic Search, Iosched, Picasso, Retrofit, and Storm, as detailed in Table~\ref{tb:projects}. We chose the aforementioned projects, since they (i)~are publicly available in GitHub, (ii)~belong to different application domains, (iii)~have more than a thousand stars, and (iv)~follows a pull-based development process, i.e., they accept pull requests from the community.

We extracted all pull requests from the six projects.\footnote{The extraction was performed on December 11, 2015}
To carry out this activity, we developed an extractor of pull requests using the GitHub API, which allowed us to easily extract pull requests that were closed along a project’s lifetime, with the \textcolor{black}{three} kinds of comment (as detailed in Section~\ref{sec:dsd}). 
Altogether, our study investigates 7,639 pull requests. 

\subsection{Selection procedure}
Our study relies on the following inclusion/exclusion criteria in order to 
analyze only consistent data.

\begin{itemize}
\item (Inclusion) We only consider pull requests that have been closed;
\item (Exclusion) We disregard pull requests that had no reviews because without comments we cannot figure out the reason for the rejection; and

\item	(Exclusion) We disregard pull requests that (i)~had no comments from at least one project's owner or (ii)~had only automatic comments from bots.
\end{itemize}

As reported in Table~\ref{tb:filtering_results}, the aforementioned criteria reduced the number of pull requests in our initial dataset. 
\textcolor{black}{The largest reduction occurred in Elastic Search, which reduced the number of pull requests from 6,115 to 584 (9.6\%)}.

\begin{table}[ht]
\centering
\caption{Filtering Results}
  \begin{tabular}{@{}p{1.5cm}@{}c@{}c@{}c@{}c@{}}
\hline
{\bf System}	& \multicolumn{1}{P{1.4cm}@{}}{\bf PR's {\scriptsize\tt (A)}} & \multicolumn{1}{P{1.5cm}@{}}{\bf	PR's After Filtering {\scriptsize\tt (B)}}	& \multicolumn{1}{P{1.5cm}@{}}{\bf Comments After Filtering} &	\multicolumn{1}{P{1.4cm}@{}}{\bf \% of selected PR {\scriptsize\tt (B/A)}}\\
\hline
Elastic Search &	6,115  & 	584	 & 1,374	 & 9.6\%\\
IntelliJ & 	323 & 	226	 & 603 & 	70.0\%\\
Iosched & 	72	 & 48 & 	62	 & 66.7\%\\
Picasso & 	344	 & 339	 & 509 & 	98.5\%\\
Retrofit & 	523 & 	319 & 	703	 & 61.0\%\\
Storm	 & 262	 & 206	 & 512 & 	78.6\%\\
\hline
{\bf Total}& 	{\bf 7,639}  & 	{\bf 1,722}	  &  {\bf 3,763}  & {\bf 22.5}\\
\hline 
  \end{tabular}
  \label{tb:filtering_results}
\end{table}

\begin{table*}[!t]
\centering
\caption{Percentage of rejected pull requests}
  \begin{tabular}{lccccc}
\hline
{\bf System}	& \multicolumn{1}{P{2.4cm}}{\bf Analyzed PR {\scriptsize\tt (A)}} 	& \multicolumn{1}{P{2.4cm}}{\bf Rejected PR {\scriptsize\tt (B)}}	& \multicolumn{1}{P{2.8cm}}{\bf PR rejected due to TD {\scriptsize\tt (C)}}	& \multicolumn{1}{P{2.4cm}}{\bf \% of rejection {\scriptsize\tt (B/A)}}  & \multicolumn{1}{P{2.4cm}}{\bf \% of rejection due to TD {\scriptsize\tt (C/B)}} \\
\hline
IntelliJ  & 	226 & 	211 & 	26 & 	93.4\% & 	12.3\%\\
Iosched & 	48 & 	22 & 	1 & 	45.8\% & 	4.5\%\\
Picasso & 	339 & 	112 & 	59 & 	33.0\% & 	52.7\%\\
Storm & 	206 & 	112 & 	32 & 	54.4\% & 	28.6\%\\
Retrofit & 	319 & 	103 & 	49 & 	32.3\%	 & 47.6\%\\
Elastic Search & 	584 & 	119 & 	39 & 	20.4\%	 & 30.3\%\\
\hline

{\bf Total}		& {\bf 1,722} & {\bf 679} & {\bf 206} &	{\bf 39.4\%}	& {\bf 31.1\%}\\
\hline
  \end{tabular}

  \label{tb:rejected_pr}  
\end{table*}

\begin{table*}[ht]
\centering
\caption{Results per category}
  \begin{tabular}{lcccccccc}
\hline
{\bf Type of TD} &	{\bf IntelliJ} &	{\bf Iosched} &	{\bf Picasso}&	{\bf Storm}&	{\bf Retrofit}&	\multicolumn{1}{p{1.2cm}}{\bf Elastic Search}& 	{\bf Total} &	\multicolumn{1}{p{2cm}}{\bf \% per type}\\
\hline
Build	& 0	&1&	2&	2&	0&	0&	{\bf 5}&	2.37\%\\
Design&	15&	0&	9&	19&	27&	13&	{\bf83}&	39.34\%\\
Documentation&	0&	0	&4&	3&	0&	5&	{\bf12}&	5.69\%\\
Performance&	2&	0&	12&	0&	3&	3&	{\bf20}&	9.48\%\\
Project Convention&	5&	0&	10&	4&	6&	8&	{\bf33}&	15.64\%\\
Security&	0&	0&	2&	0	&1	&0&	{\bf3}	&1.42\%\\
Tests	&4	&0	&20	&4	&12	&10	&{\bf50}	&23.70\%\\
Unclassified &	185&	21	&53	&80&	54&	80	&{\bf468} & ---	\\

\hline
{\bf Total}		& {\bf 211} & {\bf 22} & {\bf 112} &	{\bf 112}	& {\bf 103}	& {\bf 119} 	& {\bf 679} 	& {\bf ---}\\
\hline
  \end{tabular}

  \label{tb:results_per_category}  
\end{table*}

\subsection{Manual Investigation}

We have classified the 1,722 pull requests that remaining in our dataset after applying the selection procedure in one of the following seven different types of Technical Debts: {(i)~\em design}, (ii)~{\em documentation}, (iii)~{\em test}, (iv)~{\em build}, (v)~{\em project convention}, (vi)~{\em performance}, or (vii)~{\em security} debt.

The classification was made by the first author of this paper, who has more than nine years of experience working as a software engineer, in particular with Java programming language. 
We argue that these qualifications provide him the necessary background to conduct the manual classification of the comments. 

The first author manually read through pull requests as described in Section~\ref{sec:dsd}. 
While examining them, he classified each item by the nature of the debt. 
Some instances could be classified in more than one type of debt, e.g., a code with a poor implementation could be considered a {\em design} debt, but it could also be considered a {\em performance} debt since it causes excessive consumption of memory. 
Although this situation may have different interpretations depending of who is reading the comments, we defined that each pull request would have only one classification type for the sake of clarity. 
Basically, we considered the more meaningful type for each scenario. 
In total, we read and classified 1,772 pull requests from six open source projects.

\section{Results}
\label{sec:results}
This section provides answers for the four research questions of our study.

\rqi\\[-0.2cm]

Our results indicate that 679 out of 1,722 pull requests were rejected. Therefore, \ul{39.4\% of pull requests are rejected, on average}.\\

\rqii\\[-0.2cm]

Table~\ref{tb:rejected_pr} summarizes the data related to pull requests rejection. 
The last two columns show, respectively, the percentage of rejected pull requests and the percentage of rejected pull requests due to Technical Debt. 
The data of last column is the percentage of items with Technical Debt related to the rejected pull requests (instead of all ones).
Our results indicated that 206 out of the 679 rejected pull requests are due to Technical Debt. 
Therefore, \ul{30.3\% of the rejected pull requests are due to Technical Debt.}

It is worth noting that
Picasso is the project with the highest percentage of pull requests rejected due to Technical Debt reasons (52.7\%). Retrofit and Elastic Search immediately follows with respectively 47.6\% and 32.8\%. These high percentages suggest that Technical Debt is a very strong criterion for rejecting a pull request.\\

\rqiii\\[-0.2cm]

As reported in Table~\ref{tb:results_per_category}, we found 206 pull requests rejected due to Technical Debt. These pull requests can be classified in one of the following seven types: {\em design}, {\em documentation}, {\em test}, {\em build}, {\em project convention}, {\em performance}, or {\em security} debt.  The last column shows the percentage of each type of Technical Debt found in our study. The data is normalized, presenting the percentages of the different types rather than the raw numbers. For example, if a project has 100 pull requests that were rejected due to Technical Debt and 15 are due to design debt, then the percentage of {\em design} debt is 15\%.
\ul{Our results indicate that the most frequently Technical Debt encountered was \textit{design} (39.34\%), followed by
\textit{test} (23.70\%), 
\textit{project convention} (15.64\%), 
\textit{performance} (9.48\%), 
\textit{documentation} (5.69\%),
\textit{build} (2.37\%), and
\textit{security} (1.42\%)}.\\

{\noindent \em Design Debt}: It refers to problems such as architecture violations, use of bad programming practices, violations of good object-oriented design principles, misuse of design-pattern, misplaced code, lack of abstraction, poor implementation, and workarounds or temporary solutions. {This debt represents 39.34\% of the rejection of pull requests}. From our dataset, we can highlight the following examples:

\frase{These X* classes tend to be very painful to maintain when doing Lucene upgrades.}{Elastic Search}

\frase{Wouldn't it be better to catch and suppress FileNotFoundExceptions in rmpath itself? The ultimate goal of the function is to ensure that path is deleted, so it's fine to suppress that error.}{Storm}

{\em Design debt} was the most frequently Technical Debt encountered, although not in all projects. In Picasso, for example, the most common type was related to {\em test} debt. This may suggest that the major concerns of each team is different, as a group has a greater concern for correct use of best practices, others emphasize more well-tested systems.\\

{\noindent \em Documentation Debt}: It refers to problems found in the project documentation and can be identified by looking for missing, inadequate, or incomplete documentation of any type. It can also be related to architectural documents, code comments, Javadoc, release notes, etc. Besides, a poor description of the pull request reason is also considered a documentation problem. 
{This debt was responsible for 5.69\% of the rejection of pull requests}. We can highlight the following examples:

\frase{Closing this in favor of documenting indices.get\_gield\_mapping as `fields`, this to resemble to query string option present on several API's.}{Elastic Search}

\frase{Documentation tends to become out of sync with the code, and it can become a maintenance nightmare. I'd rather we did this as a wiki page with links into the code, ala what I started doing here: https://github.com/nathanmarz/storm /wiki/Implementation-docs}{Storm}

{\noindent \em Test Debt}: It specifies the need for implementation or improvement of the submitted tests. 
{This debt was responsible for 23.70\% of the rejection of pull requests}. We can highlight the following examples:

\frase{Do you think you'd be able to add a test for this behavior?}{Picasso}

{\noindent \em Build Debt}: It refers to any issue that could compromise build or deployment. Changes that make the build or deploy harder, consuming more time/processing unnecessarily, or hindering the deploy in some way. For example, changes in the build process can make it use more dependencies than necessary. 
{This debt was responsible for 2.37\% of the rejection of pull requests}. From our dataset, we can highlight the following examples:

\frase{[...]~looks like you're having some weird quirks on your builds~[...]}{Iosched}

{\noindent \em Performance Debt}: It refers to any part of code that can delay or hinder the system’s performance. For example, it may include any piece of code that heavily use the network or increases the use of memory heap.
 
{This debt was responsible for 9.48\% of the rejection of pull requests}. We can highlight the following examples:

\frase{We are not going to buffer every download. This will destroy the heap.}{Picasso}

\frase{This should use double-checked locking. You're optimizing for the worst case instead of the best case.}{Picasso}

{\noindent \em Project Convention Debt}: 
It deals with the format and style conventions adopted by the project. For example, names of variables, method size, and indentation. It is worth noting that this debt was initially comprehended as {\em design}. However, during our classification process, we noted that this scenario is very important in some projects. Therefore, we considered these issues as a new type. {It was responsible for 15.64\% of the rejection of pull requests}. We can highlight the following examples:

\frase{Sorry, I'm not able to accept a pull request that renames all fields in a class so that they no longer match the IntelliJ IDEA coding style. Please make the minimum amount of changes required for your refactoring and follow the style of the codebase you're contributing to.}{IntelliJ}

\begin{table*}[ht]
	\centering
	\caption{Total discussion per type of Technical Debt and per system}
	\begin{tabular}{p{2.8cm}ccccccP{1cm}P{1.4cm}P{1.5cm}}
\hline
		& \multicolumn{6}{c}{\bf \# of comments} &&& \\
		{\bf \mbox{Type~of TD}} &	{\bf IntelliJ} &	{\bf Iosched} &	{\bf Picasso}&	{\bf Storm}&	{\bf Retrofit}&	\multicolumn{1}{p{1.2cm}}{\bf Elastic Search}& 	{\bf Total per type {\scriptsize\tt(A)}} & 	{\bf Total of PR {\scriptsize\tt(B)}} &  {\bf Average {\scriptsize\tt (A/B)}} \\
	\hline
		{ Build} & \textcolor{gray}{\scriptsize 0} &	3 &	4 &	4 &	\textcolor{gray}{\scriptsize 0}	& \textcolor{gray}{\scriptsize 0}	& {\bf 11} & {\bf 5} & {\bf 2.2}  \\
		
		{ Design} &	37 &	\textcolor{gray}{\scriptsize 0} &	22 &	34 &	54 &	52 &	{\bf 199} & {\bf 83} & {\bf 2.4}   \\
		
		{ Documentation} &	\textcolor{gray}{\scriptsize 0} &	\textcolor{gray}{\scriptsize 0}	& 10 &	10 &	\textcolor{gray}{\scriptsize 0} &	12	& {\bf 32} & {\bf 12} & {\bf 2.7}  \\
		
		{ Performance}	& 12 &	\textcolor{gray}{\scriptsize 0}	& 40 &	\textcolor{gray}{\scriptsize 0} &	6 &	12	& {\bf 70} & {\bf 20} & {\bf 3.5} \\
		
		{ Project \mbox{Convention}} & 14 &	\textcolor{gray}{\scriptsize 0}   & 35 &	7 &	54 &	24   & {\bf 134} & {\bf 33} & {\bf 4.1}  \\
		
		{ Security} &	\textcolor{gray}{\scriptsize 0} &	\textcolor{gray}{\scriptsize 0}	& 8 &	\textcolor{gray}{\scriptsize 0} &	14 &	\textcolor{gray}{\scriptsize 0} &	{\bf 22} & {\bf 3} & {\bf 7.3}  \\
		
		{ Tests} &	15 &	\textcolor{gray}{\scriptsize 0} &	79 &	8 &	30 &	44 &	{\bf 176} & {\bf 50} & {\bf 3.5}   \\
		\hline
		{ Total per system} &	{\bf 78} & {\bf 3} &	{\bf 198} &	 {\bf 63} &	{\bf 158} & {\bf 144} &	{\bf 644} & {\bf 206} & {\bf 3.1} \\
		\hline
	\end{tabular}
	
	\label{tb:results_of_comments}  
\end{table*}

\begin{figure*}
	\begin{center}
		\vspace{5pt}
		\includegraphics[width=0.98\textwidth]{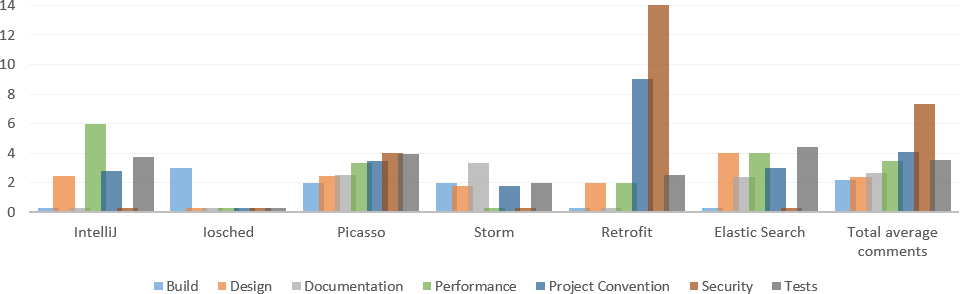}
		\caption{Average discussion per type of Technical Debt and per system}
		\label{fig:average_comments}
	\end{center}
	\vspace{-5pt}
\end{figure*}

{\noindent \em Security Debt}: It refers to a particular solution that compromises the security of the system, e.g., introducing a particular breach of security. 
{This debt was responsible for 1.42\% of the rejection of pull requests}. We can highlight the following example:

\frase{This class should be package-private. The constructor and factory method are package-private, and it's not safe for general use.}{Retrofit}

We did not classify the pull requests that did not comply with any of the defined types ({\em unclassified}).
These pull requests were rejected due to other reasons than Technical Debt, e.g., they contained bugs or compilation errors, they did not attend a business requirement, they had not performed rebase, or they had not signed contributor license agreement.\\

\rqiv\\[-0.2cm]

Last but not least, we investigated whether the type of Technical Debt influences on the size of a discussion.
Table~\ref{tb:results_of_comments} reports the total number of comments per type of Technical Debt
and also per system. For instance, {\em build} debt had a total of 11 comments in 5 pull requests,
which results in 2.2 comments per pull request on average. 
Therefore, \ul{our results indicated that the \textit{security} and \textit{project convention} debts had the highest number of comments per pull request on average.} 
While the overall average was 3.1 comments per pull request,
{\em security} and {\em project convention} debts had on average 7.3 and 4.1 comments respectively.

As a complementary analysis, 
Figure~\ref{fig:average_comments}
illustrates the average of comments by pull request per type of Technical Debt and also per system.  
For instance,
while the {\em security} debt had
the highest overall average of comments per pull request,
particularly for IntelliJ, the highest average was the {\em performance} debt having 12 comments in 2 pull requests, which results in 6.0 comments per pull request on average.

We state that the {\em security} debt achieves the highest overall average, i.e., 7.3 comments per pull request.
We conjecture that security is a very delicate topic because sometimes we must choose between increasing system security at the expense of performance, for example.
Besides {\em security}, {\em project convention} and {\em test} debts are types that 
apparently do not have such a consensus.
For example, software development conventions are not always very clear
and also vary from project to project.

{\em Test debt} also generates a considerable number of discussions.
Because tests are very important for a system, this is somewhat expected. Hetzel \cite{19} stated that it is difficult to say to what extent a system is completely covered by tests, which naturally triggers a lot of discussion.
On the other hand, {\em design debt} had one of the lowest overall average (2.4~comments per pull request),which was unexpected since such finding contrasts to recent empirical study that reported that design is a topic that causes much debate \cite{18}.
After an investigation, we conjecture that
since software design practices and patterns are well spread in the community, 
there are no longer much discussion on this~topic.

\section{Threats to Validity}
\label{sec:threats}
Our classification of pull request is manual because comments are written in natural language and therefore difficult to analyze by a machine. 
Like any human task, it is propitious to personal subjectivity. To reduce this threat, the main author double checked all classifications.

Our dataset relies on six open source projects. The projects are written in Java and publicly available for replication. We also chose projects from different domains to minimize external validity. However, as usual in empirical studies in software engineering, the results obtained may not be generalized to other projects, e.g., implemented in other languages.\\

\section{Conclusion and Future Work}
\label{sec:conclusion}
Technical Debt is used to denote shortcuts and workarounds employed in software projects, which can impact the maintainability of the project and even hinder the development if not properly addressed. This paper explores identification and classification of Technical Debt introduced by the developers through pull-based development. We investigated pull requests of six open source projects namely IntelliJ, Elastic Search, Iosched, Picasso, Retrofit, and Storm. These projects are considered renowned and they have rigorous review processes. 

We find that Technical Debt can be classified into seven types namely {\em design}, {\em documentation}, {\em test}, {\em build}, {\em project convention}, {\em performance}, and {\em security} debts.  
More important, we provide real examples of each type and the basis for its classification.
%
{Our results indicate that the most frequent Technical Debt is
\textit{design} (39.34\%), followed by
\textit{test} (23.70\%), 
\textit{project convention} (15.64\%), 
\textit{performance} (9.48\%), 
\textit{documentation} (5.69\%),
\textit{build} (2.37\%), and
\textit{security} (1.42\%)}.
We also note that type of Technical Debt influences on the size of a discussion, i.e., {\em security} and {\em project convention} debts spur more comments prior to be rejected than the other types.
As further work, we look forward to to identify how the type of Technical Debt associated with rejection influences the time of the problem solving. For this, we plan to examine pull requests that were rejected and were successfully resubmitted afterwards.

We made our dataset  publicly available.\footnote{\url{https://github.com/aserg-ufmg/sbsi2016-data/blob/master/Code\%20Review.zip}} 
We expect this dataset inspires future research in the area, e.g., using natural language processing techniques.

\section*{Acknowledgments}
Our research is supported by FAPEMIG, CNPq, Chemtech, and Siemens.\\

\bibliographystyle{plain}
\bibliography{terracopy}

\end{document}